\documentclass[twocolumn,tighten]{aastex63}
\shorttitle{Sample article}
\shortauthors{Boldrini et al.}
\graphicspath{{./}{figures/}}
\begin{document}

\title{Flattening of dark matter cusps during mergers: model of M31}

\correspondingauthor{Pierre Boldrini}
\email{boldrini@iap.fr}

\author{Pierre Boldrini}
\affiliation{Sorbonne Universit\'e, CNRS, UMR 7095, Institut d'Astrophysique de Paris, 98 bis bd Arago, 75014 Paris, France}

\author{Roya Mohayaee}
\affiliation{Sorbonne Universit\'e, CNRS, UMR 7095, Institut d'Astrophysique de Paris, 98 bis bd Arago, 75014 Paris, France}

\author{Joe Silk}
\affiliation{Sorbonne Universit\'e, CNRS, UMR 7095, Institut d'Astrophysique de Paris, 98 bis bd Arago, 75014 Paris, France}
\affiliation{Department of Physics and Astronomy, The Johns Hopkins University, Baltimore MD 21218, USA}
\affiliation{Beecroft Institute for Particle Astrophysics and Cosmology, Department of Physics, University of Oxford, Oxford OX1 3RH, UK}



\begin{abstract}

We run high resolution fully GPU N-body simulations to demonstrate that the dark matter distribution in M31 is well-fitted by a core-like profile. Rich observational data especially on the giant stellar stream provides stringent constraints on the initial conditions of our simulations. We demonstrate that accretion of a satellite on a highly eccentric orbit heats up the central parts of M31, causes an outward migration of dark matter particles, flattens the central cusp over more than a decade in scale and generates a new model-independent dark matter profile that is well-fitted by a core. Our results imply that core-like central profiles could be a common feature of massive galaxies that have been initially cuspy but have accreted satellites on nearly radial orbits.

%\textcolor{red}{

\end{abstract}



\section{Introduction}

Due to its proximity, the Andromeda galaxy (M31) provides a wealth of high precision observational data for understanding the history of M31 and of the Local Group, and more generally, galaxy formation models in a cold dark matter-dominated universe. M31 exhibits challenging features on different scales ranging from the double nucleus at its centre on a scale of a few parsecs \citep{1995AJ....110..628T,1999ApJ...522..772K} to the giant stellar stream (GSS) in its outskirts which extends to tens of kiloparsecs \citep{2001Natur.412...49I}.  

It is widely believed that the phase structure of M31, namely its GSS and its shell-like features, result from the accretion of  a satellite galaxy \citep{2004MNRAS.351..117I,2006AJ....131.1436F,2006MNRAS.366.1012F,2007MNRAS.380...15F}.  The mass, radial velocity and distance to the stream are observed with good accuracy. The accretion scenario has been simulated with high resolution and the high quality data on the stream and the shell-like features have been used to strongly limit the initial parameter space. In previous work, we ran high resolution simulations of a live M31 and an infalling satellite, and showed that in the cosmological scenario, the infalling satellite traces a highly eccentric orbit after reaching its turn-around radius, and then  falls towards the centre of M31 \citep{2014MNRAS.442..160S}. The satellite was disrupted by M31, forming the GSS, and its subsequent passages through the centre of M31 led to the formation of the shell-like features that we observe today. Our work showed that the infalling satellite was dark matter-rich and in the same plane as most of the present-day satellites of M31. Also, minor merger scenarios, as discussed above, have been successful in explaining the phase structures of M31. Major merger scenarios have also been proposed in the past \citep{2010ApJ...725..542H}, reviewed by \citep{2018MNRAS.475.2754H}, and received recent attention,  seeming to well fit the kinematic data \citep[e.g.][]{2018NatAs...2..737D}.

The density profiles of galaxy halos provide important constraints on the nature of DM. In a cold dark matter-dominated Universe, the haloes have profiles that diverge at the centres, {i.e.}\ cuspy profiles \citep{1997ApJ...490..493N,1997ApJ...477L...9F,1998ApJ...499L...5M,2010MNRAS.402...21N}. In contrast, measurements of galaxy rotation curves and dynamical models of dSph galaxies are often claimed to require shallower slopes that are consistent with a central density core \citep[e.g.][ and references therein]{1994Natur.370..629M,1994ApJ...427L...1F,2008ApJ...681L..13B,2011ApJ...742...20W,2012MNRAS.419..184A,2012ApJ...754L..39A,2014ApJ...789...63A}. However, it has been  pointed out that disequilibrium, inclination and non-circular motions can mimic the presence of cores in galaxies \citep{2006MNRAS.373.1117H,2016MNRAS.462.3628R,2019MNRAS.482..821O}. The core profiles in dwarf galaxies \citep{2015MNRAS.452.3650O,2020MNRAS.495...58S} generally present a challenge to the present model of cosmology ($\mathrm{\Lambda CDM}$). This discrepancy has become known as the core-cusp problem. In order to reconcile observations and simulations, many mechanisms involving baryons have been proposed, transforming cusps into cores via changes in the gravitational potential caused by stellar feedback redistributing gas clouds, and generating bulk motions and galactic winds \citep[e.g.][]{1996ApJ...462..563N,2001ApJ...560..636E,2010ApJ...725.1707G,2011ApJ...736L...2O,2012MNRAS.421.3464P,2013MNRAS.429.3068T,2019MNRAS.488.2387B}. The existence of shallower than NFW density profiles in low-mass galaxies has sometimes been interpreted as the manifestation of a new DM-specific  feature, such as DM self-interactions. The multiple scattering events between DM particles can result in the formation of constant density cores by removing particles from the centres of haloes \citep[e.g.][]{2012MNRAS.423.3740V,2013MNRAS.431L..20Z,2013MNRAS.430...81R,2018MNRAS.476L..20R}.  Alternatively, in a recent study, we have shown that DM candidates in the form of primordial black holes can induce a cusp-to-core transition in low-mass dwarf galaxies via dynamical friction by DM particles \citep{2020MNRAS.492.5218B}. Yet another mechanism  could be due to  sinking massive objects, such as subhalos, gas clumps or globular clusters, that transfer energy and angular momentum to the DM field via dynamical friction, creating a DM core from an initially cuspy density distribution. The DM density profile of the galaxy flattens due to  heating via dynamical friction \citep{2001ApJ...560..636E,2004PhRvL..93b1301M,2006ApJ...649..591T,2008ApJ...685L.105R,2009ApJ...691.1300J,2010ApJ...725.1707G,2011MNRAS.418.2527I}.

In this work, we describe  another consequence of  accretion of a satellite on a highly elongated orbit   in a $\Lambda$CDM Universe, and show that the density profile of M31 could have been strongly influenced by this mechanism. Such an orbit is motivated, and indeed required,  by dynamical modelling of features in M31. We demonstrate that the passage of the satellite, which is at the origin of the GSS, the shells and the warp of the disc of M31, should have also caused a cusp-to-core transition at the centre of the DM distribution in M31.  A more general consequence of our work is that accreting satellites on highly eccentric orbits can induce a cusp-to-core transition in CDM haloes. These cores are a common feature of many DM haloes that have been initially cuspy but have accreted subhaloes on highly eccentric orbits. In most cosmological simulations, we expect to see this effect only for DM haloes with masses higher than $10^{12}$ M$_{\sun}$ by assuming that they have sufficient resolution to determine the shapes of DM density profiles \citep{2004PhRvL..93b1301M,2007MNRAS.377L...5G,2015MNRAS.451.1177L}. The DM density profile of the Aquarius simulation revealed that density profiles become shallower inwards down to the innermost resolved radius \citep{2010MNRAS.402...21N}. This slight deviation from the NFW model could be evidence for the proposed cusp-to-core mechanism \citep{2001ApJ...560..636E,2004PhRvL..93b1301M}. Indeed, most haloes have suffered multiple subhalo mergers, especially at early epochs (z$\sim$ 2-3) \citep{2008MNRAS.388.1792N}. However, the cusp can regenerate itself. As such, the cuspy profiles are more common at recent epochs, possibly explaining the presence of transient cores (see \cite{2015MNRAS.449L..90L,2003MNRAS.341..326D,2020MNRAS.492.3169B})

In this work, we perform high resolution $N$-body simulations with GPUs, which allow parsec resolution, to study the effect of the accretion of a satellite on the central density profile of the DM halo of M31. The initial conditions of our simulations are determined by  observations of the mass, density profile, radial velocities  and distances of the GSS which provide high precision tests of our model. We consider infalling galaxy scenarios with different halo-to-stellar mass ratios ranging from 0 \citep{2007MNRAS.380...15F} to 20  \citep{2014MNRAS.442..160S} and to 100 in three new models. By analysing the density profile of the halo of M31, we see that the initial cuspy profile becomes shallower as the  DM particles are heated, during the passages of the infalling galaxy, and some DM particles migrate outwards from the central region of M31 in all our five models. As the flattening of the cusp occurs for all of these models, we propose that the central DM core-like profile of M31 is model-independent.

The paper is organized as follows. In Section 2 we present a brief summary of observational data. Section 3 provides a description of the $N$-body modelling of M31 
and its satellite, along with details of our numerical simulations. In Section 4, we present the results from simulations and discuss the implications for the cusp-core problem. Section 5 presents our conclusions.

\begin{deluxetable*}{ccccccccc}
\tablenum{1}
\tablecaption{Simulation parameters}
\tablehead{&&&&The host galaxy}
\startdata
    & Component & Profile & $a$ & $r_{200}$ &  Mass & ($x_0$,$y_0$,$z_0$) & ($v_{x0}$,$v_{y0}$,$v_{z0}$)\\
    & & & [kpc] & [kpc] & [$10^{10}$ M$_{\sun}$] & [kpc] & [km.$s^{-1}$]\\
    \hline
   M31 &DM halo & NFW & 7.63 & 195 & 88 & 0 & 0 \\
   &Bulge & Hernquist & 0.61 & - & 3.24  & 0 & 0 \\
   &Disk & Exponential & $R_{\mathrm{d}}=$ 5.4 & - & 3.66 & 0 & 0\\
   && disk & $z_{\mathrm{d}}=$ 0.6 & - & - & - & -\\
   \hline
   \\
   &&&&The infalling\\
   &&&&satellite
   \\
   Scenario &\\
   \hline
    \cite{2007MNRAS.380...15F} & \\
  $\left(M_{\mathrm{DM}}/M_{*}\right)_{\mathrm{sat}}=0$ & Stars & Plummer & 1.03 & - & 0.22 & (-34.75,19.37,-13.99) & (67.34,-26.12,13.5)\\
    \hline
    \cite{2014MNRAS.442..160S}\\
   $\left(M_{\mathrm{DM}}/M_{*}\right)_{\mathrm{sat}}=20$ & DM halo & Hernquist & 12.5 & 20 & 4.18 & (-84.41,152.47,-97.08) & 0 \\
   & Stars & Plummer & 1.03 & - & 0.22 & (-84.41,152.47,-97.08) & 0\\
    \hline
    Satellite A \\
    $\left(M_{\mathrm{DM}}/M_{*}\right)_{\mathrm{sat}}=100$ & DM halo & Hernquist & 25 & 41 & 22 & (-84.41,152.47,-97.08) & 0 \\
   & Stars & Plummer & 1.03 & - & 0.22 & (-84.41,152.47,-97.08) & 0\\
    \hline
    Satellite B \\
    $\left(M_{\mathrm{DM}}/M_{*}\right)_{\mathrm{sat}}=100$ & DM halo & Hernquist & 25 & 123 & 22 & (-84.41,152.47,-97.08) & 0 \\
   & Stars & Plummer & 1.03 & - & 0.22 & (-84.41,152.47,-97.08) & 0\\
    \hline
    Satellite C \\
    $\left(M_{\mathrm{DM}}/M_{*}\right)_{\mathrm{sat}}=100$ & DM halo & Hernquist & 25 & 41 & 22 & (-168.82,304.94,-194.16) & 0 \\
   & Stars & Plummer & 1.03 & - & 0.22 & (-168.82,304.94,-194.16) & 0\\
    \hline
\enddata
\tablecomments{From left to right, the columns give for each component: the initial density profile; the scale length; the virial radius; the mass; the initial positions in a reference frame centered on M31 with the x-axis pointing east, the y-axis pointing north and the z-axis corresponding to the line-of-sight direction; the velocities in this reference frame. We set the particle resolution of all the live objects to $4.4\times10^{4}$ M$_{\sun}$ and the gravitational softening length to 10 pc for all components.}
\label{tab1}
\end{deluxetable*}

\section{Observation: Giant stellar stream and shell-like features of M31}

Phase structures, such as streams and shells, are ubiquitous in the Universe where galaxies are interacting through gravity and frequently undergo mergers \citep{1980Natur.285..643M,1983ApJ...274..534M,1988ApJ...331..682H,1989ApJ...342....1H}. Our nearest neighbour M31 provides an excellent laboratory for study of such tidal features. M31 exhibits a giant stellar stream (GSS) and shells and rings  \citep[]{2001Natur.412...49I,2005ApJ...634..287I,2006Natur.443..832B,2007ApJ...671.1591I,2002AJ....124.1452F,2003A&A...405..867B,2004ApJ...612L.117Z,2006ApJ...652..323B,2008AJ....135.1998R,2009Natur.461...66M}. 
The main focus of this paper is the GSS, which provides stringent constraints on our models. The GSS is a faint stellar tail with a mass of $2.4\times10^8$ M$_{\sun}$ (corresponding to a luminosity of $3.4\times10^7$ L$_{\sun}$ and a mass to light ratio of 7) and extends to a projected radius of about 68 kpc on the sky  \citep{2001Natur.412...49I,2006MNRAS.366.1012F} and is further accompanied by two stellar shells \citep{2002AJ....124.1452F,2007MNRAS.380...15F,2010ApJ...708.1168T,2012MNRAS.423.3134F, 2005ApJ...622L.109F,2008AJ....135.1998R}. The similarity in the stars of the GSS and the shells has been the main reason why many models have taken these features to be of similar origin \citep{2004MNRAS.351..117I,2006AJ....131.1436F,2007MNRAS.380...15F}.

An empirical minor merger scenario in which a satellite galaxy with no dark matter falls from a very close distance to M31 very recently (less than 1 Gyr ago) has been extensively used in the literature to model the GSS and the shells \citep{2006MNRAS.366.1012F,2007MNRAS.380...15F}. This model however suffered from simplifications. Firstly, the dynamical friction effect was ignored as M31 was not treated as a live galaxy, secondly, the satellite had no DM in spite of most dwarfs being DM-rich, and thirdly, the satellite started its infall from a very close-in  radius. The latter is puzzling as the satellite would have been disrupted on such a highly radial orbit (see \citep{2014MNRAS.442..160S} for full details).

In recent work, we proposed a new model for the origin of the GSS, the shells and also the warped structure of the M31 disc itself \citep{2014MNRAS.442..160S}. In our model, a dark-matter-rich satellite is accreted and falls from its first turnaround radius on an eccentric orbit onto M31. The best agreement with the observational data is obtained when the satellite falls in  about 2 Gyr ago and remains   on the same plane that presently contains many of the dwarfs of M31 \citep{2013ApJ...766..120C,2013Natur.493...62I}. Unlike the previous model, in which the satellite made no impact on the disc as it was DM-poor, the disc of M31 is perturbed by the infall of the massive satellite in our model and becomes warped.

In this paper, we use this cosmologically-motivated scenario to set up our simulations. To show that our results are universal and hold for a wide range of initial conditions, we also run simulations for the model proposed by \cite{2007MNRAS.380...15F} in which the satellite is DM-poor and starts its infall from a much closer distance to the centre of M31. In addition, we also ran simulations for a higher halo-to-stellar mass ratio $M_{\mathrm{DM}}/M_{*}$ of 100, which seems to be favoured by some models of galaxy formation and evolution \citep[e.g.][]{2011MNRAS.413..101G,2015MNRAS.451.2663H,2014arXiv1412.2748S}. For the latter, we run three simulations for satellites with different DM density profiles. Our simulations are summarized in Table~\ref{tab1}.

A major improvement to our previous work \citep{2014MNRAS.442..160S} is achieved here because we have used a fully GPU-scaled code and gain in mass resolution by a factor of 100. This   allows us to study the impact of the infalling satellite not just on the outer parts of M31 but also on the DM distribution at its centre. The rich observational data on the GSS and shells of M31 provide rather demanding tests of our model. In the following section we discuss the details of our numerical simulations.

\begin{figure}
\centering
\includegraphics[width=0.47\textwidth]{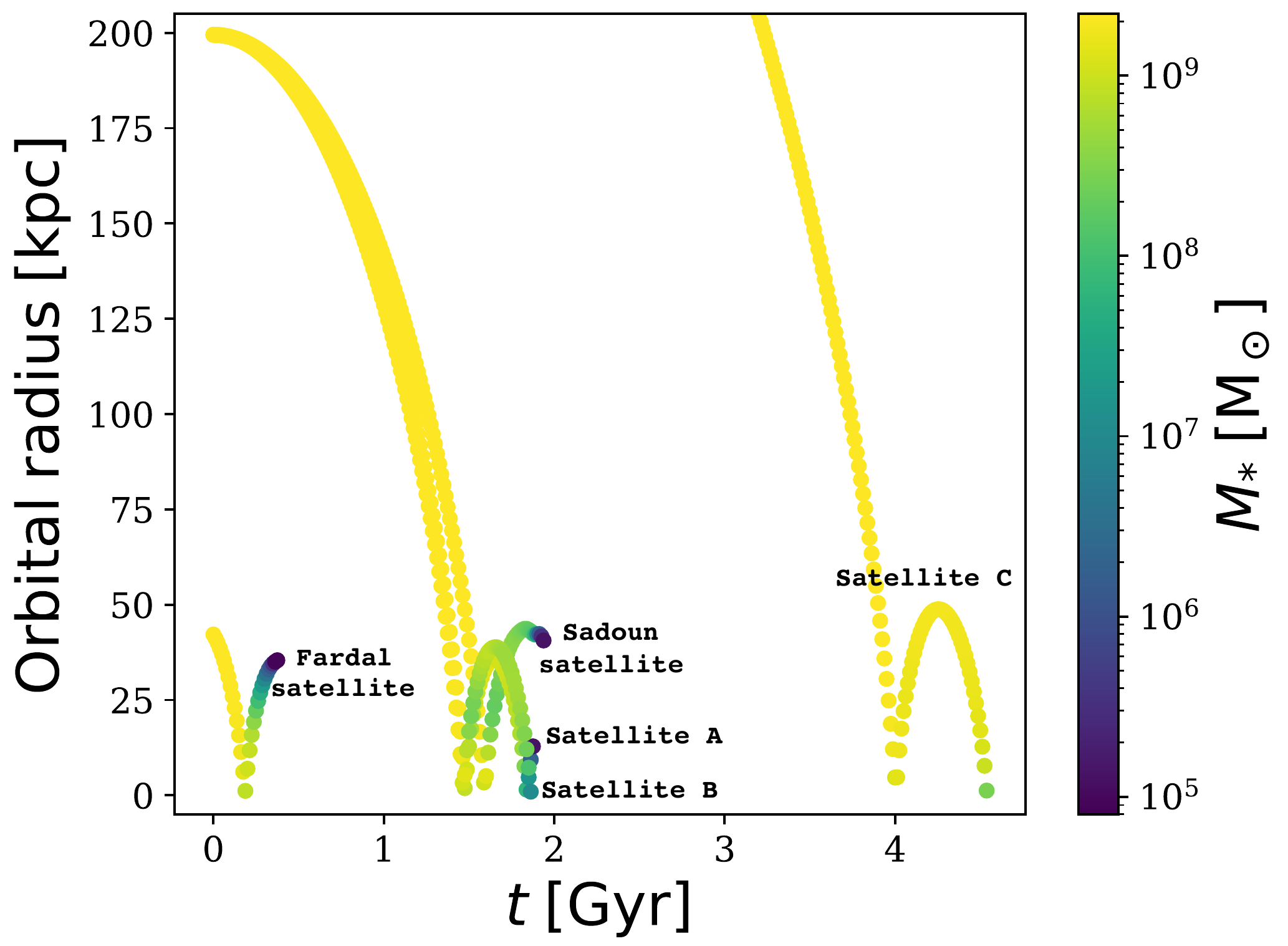}
\caption{{\it Orbital evolution of satellite:} Orbital radius as function of time in all scenarios (see Table~\ref{tab1}). The orbital radius is colour coded according to the bound stellar mass in the satellite at each time. Satellites with $\left(M_{\mathrm{DM}}/M_{*}\right)_{\mathrm{sat}}=0$, 20 and satellite A have reached their pericentre at 0.18, 1.6 and 1.47 Gyr and were completely tidally stripped after 0.37, 1.93 and 1.87 Gyr, respectively. However, models B and C exhibit the presence of a remnant of the satellite after 1.86 and 4.54 Gyr, respectively.}
\label{fgR2}
\end{figure}

%----------
\section{Simulation: High resolution fully GPU code}
%----------

The initial conditions for the M31 satellite are taken from \cite{2014MNRAS.442..160S,2007MNRAS.380...15F} and furthermore, we simulate three further models for a DM-rich satellite with $M_{\mathrm{DM}}/M_{*}=100$ (see details in Table.~\ref{tab1}). To generate our live objects, we use the initial condition code \textsc{magi} \citep{2017arXiv171208760M}. Adopting a distribution-function-based method ensures that the final realization of the galaxy is in dynamical equilibrium \citep{2017arXiv171208760M}. We perform our simulations with the high performance collisionless $N$-body code \textsc{gothic} \citep{2017NewA...52...65M}. This gravitational octree code runs entirely on GPU and is accelerated by the use of hierarchical time steps in which a group of particles has the same time step \citep{2017NewA...52...65M}. We evolve the M31 galaxy-satellite system over a few Gyr depending on the scenario. We set the particle resolution of all the live objects to $4.4\times10^{4}$ M$_{\sun}$ and the gravitational softening length to 10 pc. The softening value was estimated using the following criterion: $\epsilon\sim a/N^{1/3}$, where $N$ and $a$ are the number of particles and the scale length of the component, respectively. As all components in our simulations share the same softening, we adopt  the smaller value, which is given by the stellar component of the satellite. As in \textsc{Gadget-2} \citep{2005MNRAS.364.1105S}, the acceleration Multipole Acceptance Criterion is employed in our GPU code \citep{2017NewA...52...65M}. The time-step parameter $\eta =$ is set to 0.5.

To compare with previous simulations, we comment that \cite{2014MNRAS.442..160S} used \textsc{nbodygen} for initial conditions and \textsc{gadget-2}, here we test our models using our fully GPU code \textsc{gothic} and \textsc{magi} as the initial condition generator, which allows us to achieve 100 times higher mass resolution. This higher resolution enables us to study the impact of the satellite on the central DM density profile of M31.

%----
\section{Tests of the models with M31 observations}
%------

\begin{figure*}
\centering
\includegraphics[width=\textwidth]{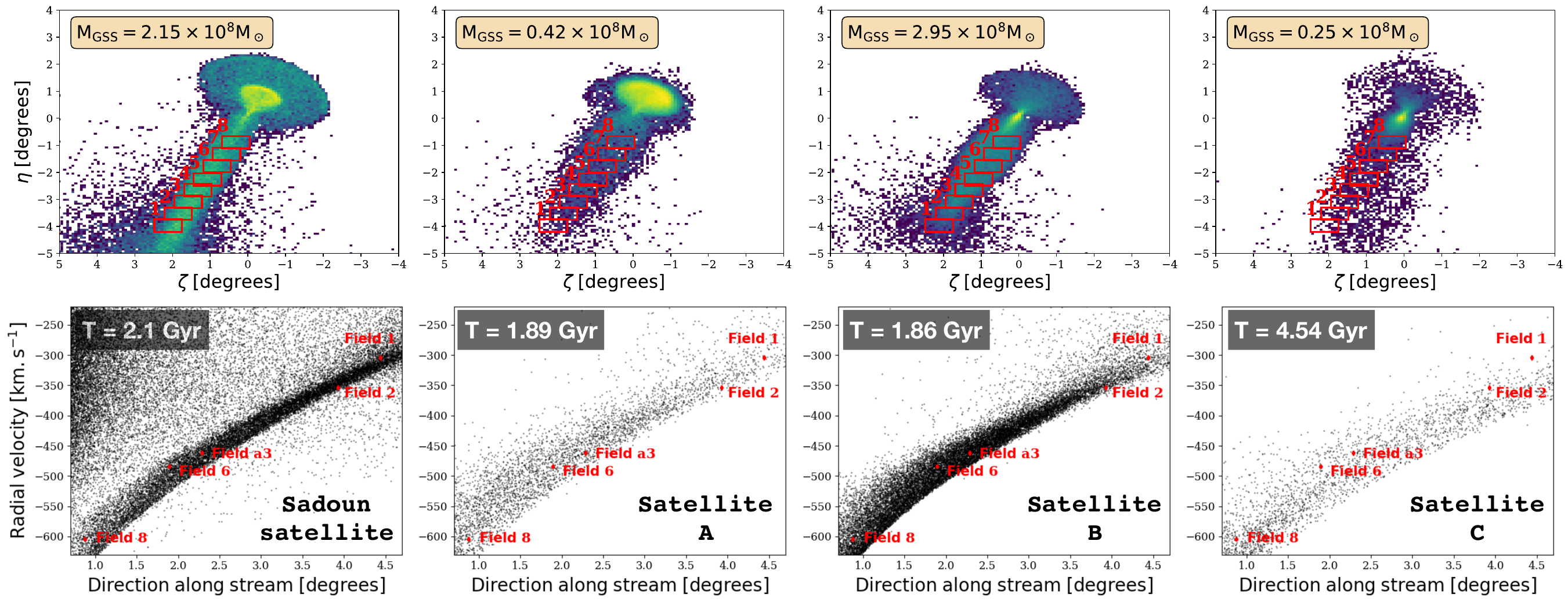}
\caption{{\it Satellite models constrained by the GSS:} Top panels: Simulated stellar density maps in standard sky coordinates corresponding to stars of the satellite at different times. We represent the observed stellar fields by black boxes \protect{\citep{2003MNRAS.343.1335M}}. We find $M_{\mathrm{GSS}}=2.15\times10^8$ M$_{\sun}$ for Sadoun model in good agreement with the value of $2.4\times10^8$ M$_{\sun}$ derived from observations with a mass-to-light ratio of 7 \citep{2001Natur.412...49I,2006MNRAS.366.1012F}. However for models with large $M/M=100$, the mass of the GSS is overestimated in model B and models A and C do not match any of the observations. Bottom panels: Simulated radial velocity of satellite particles as a function of the distance along the stream at different times when the best with observations are obtained. We represent the radial velocity measurements in five fields by red points with error-bars \citep{2004MNRAS.351..117I,2006MNRAS.366.1012F}. A good agreement with observations for the radial velocity measurement is shown for the Sadoun model and to a lesser extent for model B.}
\label{fgR100a}
\end{figure*}

We consider five scenarios for the formation of the GSS of M31. The original empirical scenario in which the satellite has no DM \citep{2007MNRAS.380...15F} and the cosmological motivated scenario in which the satellite is DM-rich \citep{2014MNRAS.442..160S} (see Table~\ref{tab1}). Many models of formation and evolution of galaxies favour a larger halo-to-stellar mass ratio $M_{\mathrm{DM}}/M_{*}$ of about 100 whereas the model proposed by \cite{2014MNRAS.442..160S} favours a lower $M_{\mathrm{DM}}/M_{*}$ of 20. Interestingly, this implies that the satellite lies below the halo-to-stellar mass relation as galaxies in this mass range should have a mean halo mass of $\simeq 2\times 10^{11}$ M$_{\sun}$ \citep{2010ApJ...717..379B,2013MNRAS.428.3121M,2015MNRAS.451.2663H}.

To conform with these observations of dwarf galaxies, we further ran three simulations with $M_{\mathrm{DM}}/M_{*}=100$ and with different density profiles and initial infall radii, but keeping $10^{9}$ M$_{\sun}$ for the stellar component as the mass of the GSS is well constrained by the observations. In these simulations, the satellite has a cuspy profile as summarized in Table~\ref{tab1}.

The orbital evolution of satellites in these five scenarios up to their passages through the centre of M31 and their subsequent disruptions is shown in Figure~\ref{fgR2}. The orbital radius is colour coded according to the bound stellar mass in the satellite at each time. We follow the iterative method of \cite{2003MNRAS.340..227B} to determine the number of bound stars over time. Since all our models are stringently constrained by the observational data on the GSS and shells of M31, we stop the simulations when the best agreements with these kinematic data are achieved. Hence, some of these mergers are constrained to be more ancient than others as clearly depicted both on Figure~\ref{fgR2} and ~\ref{fgR100a}. Satellite galaxies with $M_{\mathrm{DM}}/M_{*}=0$ and 20 have reached their pericentre at 0.18 and 1.6 Gyr and were completely tidally stripped within 0.37 and 1.93 Gyr, respectively. Satellites with $\left(M_{\mathrm{DM}}/M_{*}\right)_{\mathrm{sat}}=0$, 20 and satellite A have reached their pericentres at 0.18, 1.6 and 1.47 Gyr and were completely tidally stripped after 0.37, 1.93 and 1.87 Gyr, respectively. However, models B and C exhibit the presence of a remnant of the satellite after 1.86 and 4.54 Gyr, respectively.

In our best-fit model \cite{2014MNRAS.442..160S}, the simulation goes back to 2.1 Gyr ago and that is why we cannot exclude the cosmological turn-around of the infalling satellite as part of its real orbit. Our $r_{\mathrm{peri}}/r_{\mathrm{apo}}$ matches quite well the result of \cite{2010MNRAS.406.2312L} if we evaluate this ratio for the follow-up apocentres as is done in \cite{2010MNRAS.406.2312L} and is demonstrated for our model in Figure~\ref{fgR2}. We find that $r_{\mathrm{peri}}/r_{\mathrm{apo}}$ is about 0.1, which is exactly within the domain of values given by \cite{2010MNRAS.406.2312L}.

For models with $M_{\mathrm{DM}}/M_{*}=100$, the cases A and C (see Table~\ref{tab1}) do not match well the kinematic data. We find that the DM in the satellite prevents the stellar component from  being tidally stripped after the first pericentre passage and consequently the resulting GSS is far less massive than required by the observations. Subsequent passages will be necessary to completely strip the satellite but  once again fail to reproduce the observations (see Figure~\ref{fgR100a}). For model B in which the satellite has the same initial orbital conditions as in \cite{2014MNRAS.442..160S} but has a much larger concentration parameter, the mass of the GSS is slightly overestimated but the other data are well matched (see Figure~\ref{fgR100a}).For the Sadoun model, we find $M_{\mathrm{GSS}}=2.15\times10^8$ M$_{\sun}$ in excellent agreement with the value of $2.4\times10^8 $ M$_{\sun}$ derived from observations \citep{2001Natur.412...49I,2006MNRAS.366.1012F}.

\begin{figure*}
\centering
\includegraphics[width=\textwidth]{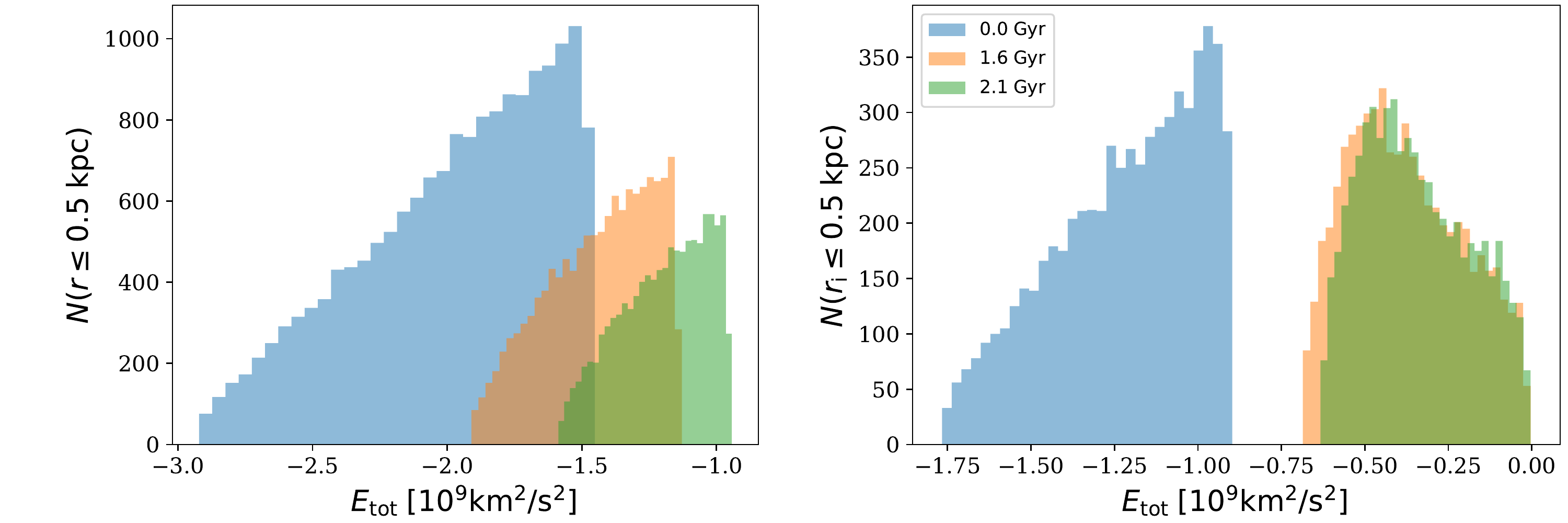}
\caption{{\it Energy transfer via dynamical friction:} In the left panel, we show the evolution of the energy distribution for DM particles of M31 within a fixed radius of 500 pc in the Sadoun scenario (see Table~\ref{tab1}). The time for the snapshots are chosen because the pericentre passage occurs at about 1.6 Gyr and the simulation is stopped at 2.1 Gyr when the best agreements with observations are achieved. The figure confirms that not only the DM particles that remain within 500 pc have gained in energy but also some of them have migrated out of this region. In the right panel, we follow the DM particles which were initially inside the 500 pc radius and show the evolution of the distribution of the energy at the same time intervals as the left panel. Hence in this figure, unlike the left panel, the number of DM particles is fixed. The histogram clearly shows that most of the particles that were initially inside the 500 pc radius, whether migrated or not, have heated up which could plausibly be a mechanism for  core formation.}
\label{fgR3}
\end{figure*}

In addition to fitting the mass of the GSS, we select our best-fit model by making a detailed comparison with other observations of the stellar stream for all the five scenarios (see Table~\ref{tab1}). Figure~\ref{fgR100a} shows the stellar density maps in standard sky coordinates corresponding to stars of the satellite. We represent the observed stream fields as solid rectangles with proper scaling. We note that the simulated stream is in good agreement with the observations regarding the morphology and spatial extent of the GSS only for the Sadoun model and to a lesser extent for model B (see Figure~\ref{fgR100a}). Furthermore, we test the five models against kinematic data. Figure~\ref{fgR100a} shows radial velocities of satellite particles as a function of the distance along the stream. We obtain good agreement with observations of the radial velocity measurement in the five fields for the Sadoun model and model B \citep{2004MNRAS.351..117I,2006MNRAS.366.1012F}. 

The small halo-to-stellar mass ratio adopted in our best-fit model \citep{2014MNRAS.442..160S} can be puzzling in the general context of formation and evolution of galaxies, which in general prefers a larger DM mass for the satellite of stellar mass in the range used in this model ($M_*=2\times10^9$ M$_{\sun}$). However, there are many incidences where a small halo-to-stellar mass ratio as such is favoured. It is widely believed that dwarfs formed in tidal tails have a low dark matter (DM) content \citep{2020NatAs...4..246G}. It has also been proposed that dwarfs, that form in high velocity dwarf mergers, are DM deficient \citep{2020ApJ...899...25S}. There are also cases of dwarf galaxies that grow passively \citep{2019ApJ...874..114C}. It has also been argued that many of the dwarfs of the Milky Way are DM-poor \citep{2020ApJ...892....3H}. These are just  a few arguments in favour of the low halo-to-stellar mass ratio used in Sadoun model. Here, we emphasize that a large halo-to-stellar mass ratio was dismissed by \cite{2014MNRAS.442..160S} because it would thicken the disk of M31 by more than is  allowed by observations. Here we have run three further simulations with $M_{\mathrm{DM}}/M_{*}=100$ and have shown that one could not satisfy all observational constraints set by the GSS as well as in the Sadoun model. Indeed, model B which uses identical initial orbital conditions as \cite{2014MNRAS.442..160S} and which the merger dates back to about 2 Gyr as in \cite{2014MNRAS.442..160S} fits well the observations but slightly overestimates  the mass of the GSS.

%--------
\section{Results: A model-independent cusp-to-core transition in M31}
%--------

\begin{figure*}
\centering
\includegraphics[width=0.9\textwidth]{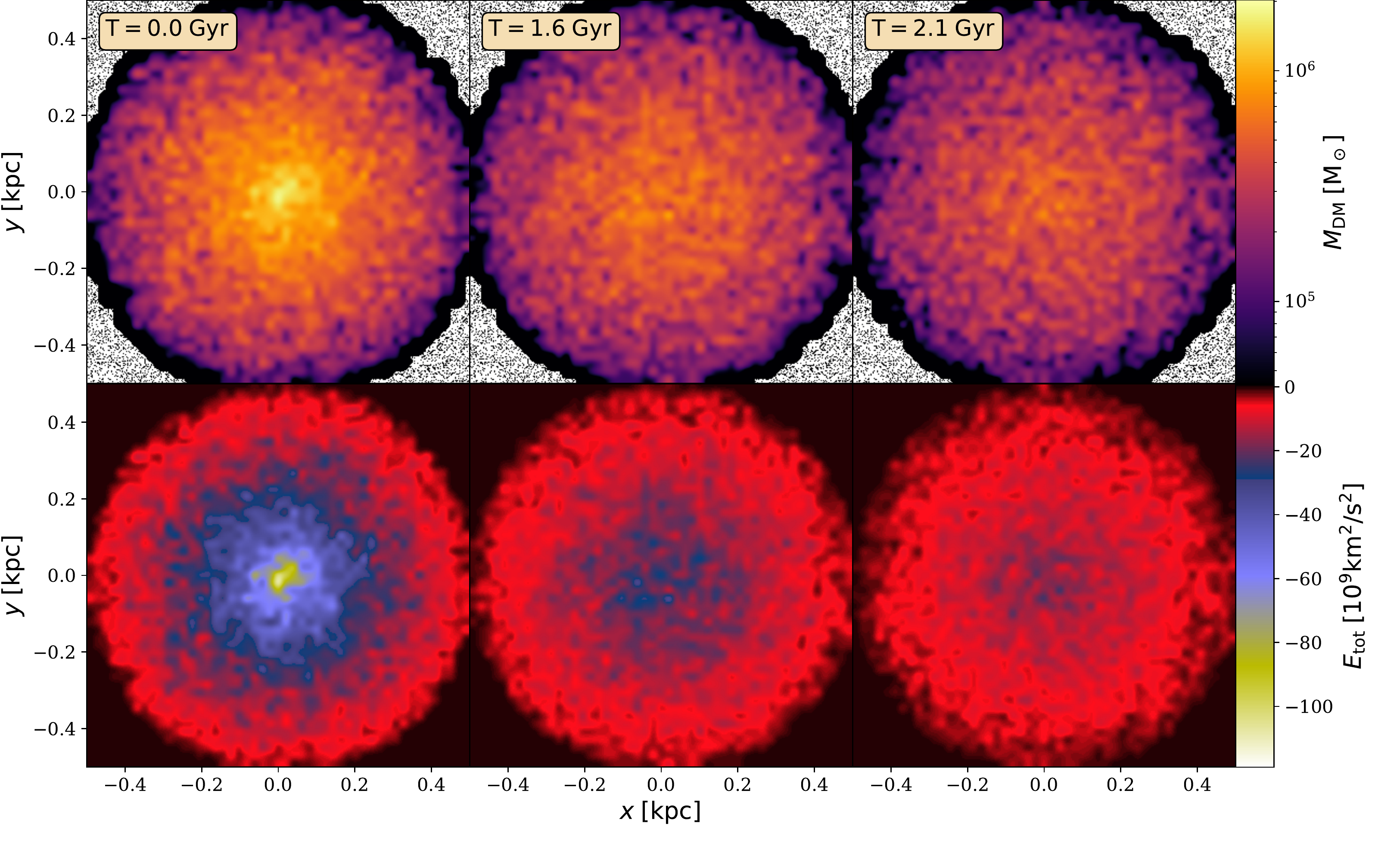}
\caption{{\it Migration and heating of DM in the central region of M31 during the passage of a satellite:} Density-weighted projected mass (top panel) and total energy maps (bottom panel) of a slice of thickness 200 pc for M31 DM particles at different times in Sadoun scenario (see Table~\ref{tab1}). Top panel shows the migration of DM particles in the central region of M31, especially after the pericentre passage of the satellite (1.6 Gyr). The bottom panel shows the heating of the DM particles that remains within the 500 pc radius central region.}
\label{fgR4}
\end{figure*}

\begin{figure*}
\centering
\includegraphics[width=\textwidth]{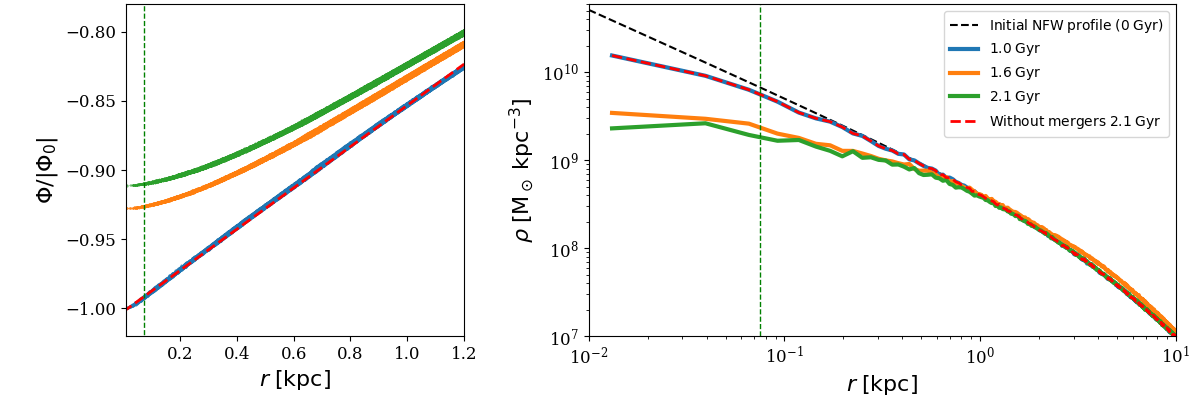}
\caption{{\it Change in the potential profile and flattening of the density profile of M31:} Profile of the DM potential of M31 normalized by the norm of its initial central value $|\Phi_0|$ at different times in Sadoun scenario (see Table~\ref{tab1}) is shown in the left panel. The spherically-averaged DM density profile of M31 in the Sadoun scenario in 26 pc thick radial shells at different times is shown in the right panel. Initially, the M31 DM halo assumes a NFW profile (black dashed curve). The convergence radius of 73 pc for the M31 DM halo is shown by the vertical dashed green line below which the simulations cannot be considered to be fully converged according to the criterion of \protect\cite{2003MNRAS.338...14P}. Beyond this region, the flattening of the cusp over almost one decade is evident. In the absence of a satellite, the DM density profile remains cusp. This scenario ensures the stability of the M31 profile against numerical effects (red dotted curve). The times for the snapshots are chosen because the pericentre passage occurs at about 1.6 Gyr and the simulation is stopped at 2.1 Gyr when the best agreements with observations are achieved. Both panels demonstrate the gradual flattening of the initial cuspy profile.}
\label{fgR5}
\end{figure*}

\subsection{Heating and migration of dark matter in M31}

In this subsection, we study the heating and migration of DM in M31 for our best-fit model \citep{2014MNRAS.442..160S}. Unlike the previous work of \cite{2014MNRAS.442..160S} which lacked resolution and computational power, our GPU simulations enable us to study the impact of the accretion of a DM-rich satellite on the DM distribution in the central regions of M31. In order to determine properly the M31 centre, we use the method of \cite{2003MNRAS.338...14P} by applying the shrinking sphere method to M31 (DM halo and the satellite), the M31 halo and the M31 bulge and have found similar centres at each time. As discussed in \cite{2003MNRAS.338...14P}, this iterative technique localised efficiently the densest region in halos, which are distorted by mergers. We have further checked our result against this method by finding particles with the lowest potential as the center of the halo in order to have a robust determination of the center. In the end, all our simulations are centered on the densest region of the M31 halo. 

As the satellite has a radial orbit, its crossings near the centre perturb the M31 halo by heating its DM particles via dynamical friction. The pericentre passage of the Sadoun satellite occurs at 1.6 Gyr. In our merger scenarios, the two DM halos interact gravitationally via dynamical friction. At the particle scale, this means that energy transfers occur between DM particles. This can be seen as a heating of DM particles of M31 from the DM particles of the satellite. Indeed, the energy and angular momentum of the satellite galaxy is consumed by dynamical friction. The  result is that  that these energy transfers are going to expand the orbits of DM particles of M31. However, at kpc scales, we expect to observe the emergence of collective effects and {\it potential fluctuations}, which erode the central density cusp of the M31 halo: the transfer of the gravitational energy from the satellite to M31 perturbs the DM distribution of the host halo.

Figure~\ref{fgR3} depicts the total energy distribution for DM particles of M31, which are within the 500 pc ({\it left panel}) and in the {\it right panel} we follow the particles that were initially within  500 pc at different times in the Sadoun scenario. As the number of particles is not conserved in the left panel of Figure~\ref{fgR3}, it confirms that some DM particles have gained in energy and have migrated outwards from the central region of M31. The right panel demonstrates that the particles that were initially within 500 pc have gained in energy.

To further confirm the mechanism of migration and heating behind the core formation, we also show density-weighted projected mass maps in Figure~\ref{fgR4}. The top panel of this figure clearly demonstrates the migration from the central region as the mass within the 500 pc radius decreases after the pericentre passage of the satellite at 1.6 Gyr. Not only the migrated DM particles but also those that remain within the central 500 pc have gained in energy through dynamical friction which slows down the satellite.

The heating and particle migration in the central region of M31 is expected to lead to the flattening of the DM density profile. In Figure ~\ref{fgR5}, we demonstrate the change in the potential (left panel) and the flattening of the cuspy initial density profile of M31 after the passage of the satellite (at 1.6 Gyr). Initially, the M31 DM halo assumes a NFW profile. We consider DM particles from both M31 and satellite haloes to determine the spherically averaged DM density profile of M31 in 26 pc thick radial shells over  time. Below a radius of 73 pc, the simulations  are affected by two-body relaxation \citep{2003MNRAS.338...14P} and beyond this region, Figure~\ref{fgR5} clearly demonstrates the flattening of the cusp over more than one decade in radius.

The flattening of the initially cuspy profile can also be seen in the change of the potential energy of M31. The profile of the potential of  M31 plotted in the left panel of Figure~\ref{fgR5} is determined by computing the gravitational potential at different radii using all DM particles. The potential at a specified position is obtained by summing all interactions with the surrounding point masses with \textsc{pNbody} \citep{2013ascl.soft02004R}.

The migration of DM particles from the central regions of M31 is expected to lead to the shallowing of the potential well, which is clearly demonstrated in Figure~\ref{fgR5}. The two panels of Figure~\ref{fgR5} together with Figures~\ref{fgR3} and ~\ref{fgR4} show clearly that energy transfer from the satellite to M31 and subsequent outward migrations of DM particles are responsible for the flattening of the density profile.

\subsection{Convergence tests: two-body relaxation, softening and mass resolution}

\begin{figure}
\centering
\includegraphics[width=0.47\textwidth]{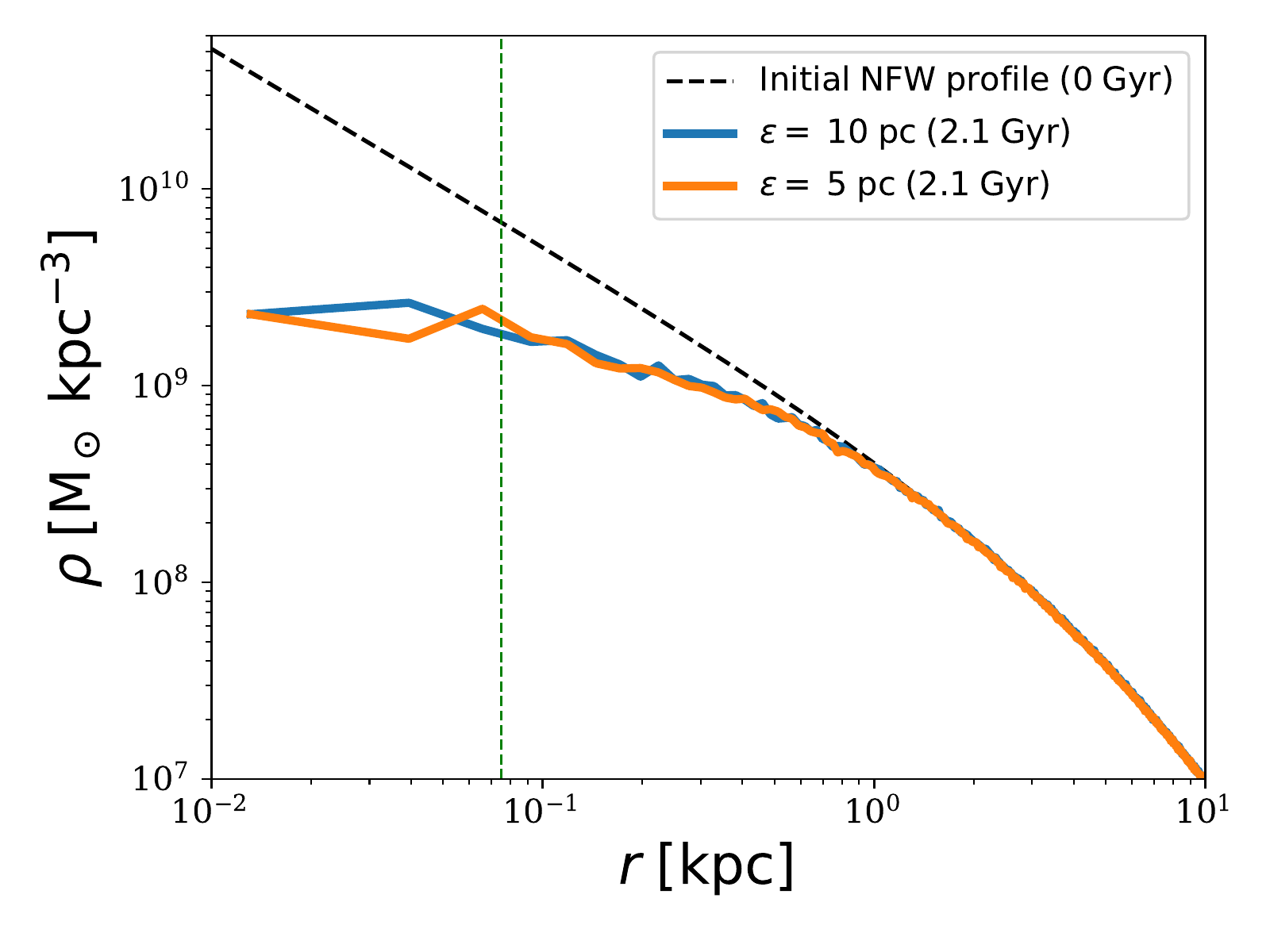}
\caption{{\it Impact of softening on DM profile of M31:} DM density profile of M31 halo for $\epsilon=10$ and 5 pc in the Sadoun scenario (see Table~\ref{tab1}). As the softening length $\epsilon$ does not affect the DM density profile, our simulation results are thus free from such numerical artifacts. The convergence radius of 73 pc for the M31 DM halo is shown by the vertical dashed green line below which the simulations cannot be considered to be fully converged according to the criterion of \protect\cite{2003MNRAS.338...14P}}
\label{app1}
\end{figure}

\begin{figure}
\centering
\includegraphics[width=0.47\textwidth]{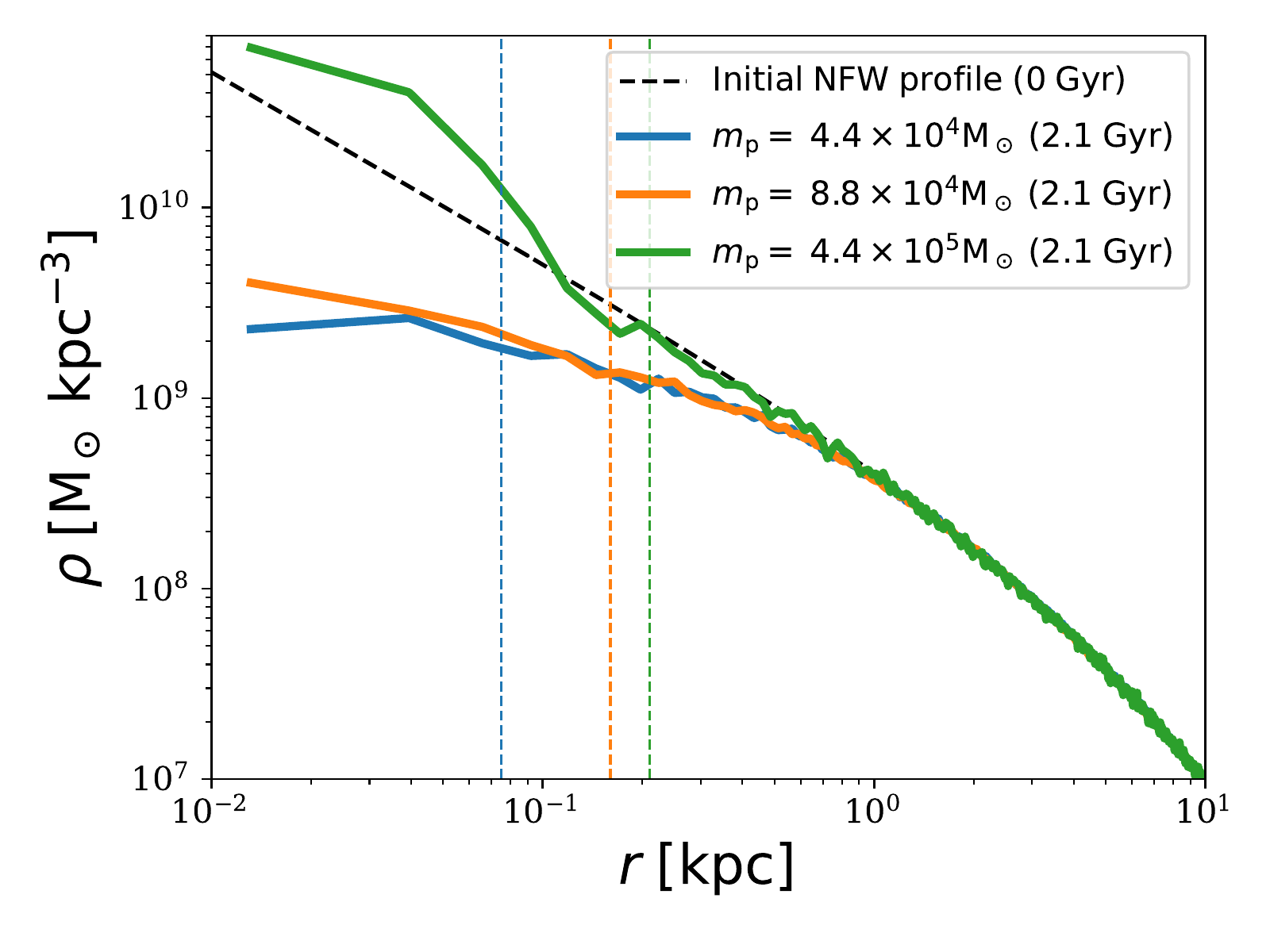}
\caption{{\it Impact of mass resolution on the DM profile of M31:} DM density profile of the M31 halo for three different mass resolutions is shown for the Sadoun scenario (see Table~\ref{tab1}). The convergence radii below which the simulations cannot be considered to be fully converged \protect\citep{2003MNRAS.338...14P} change with the mass resolution and are show by the vertical dashed lines. We see that convergence is achieved for a mass resolution smaller than $8.8\times10^4$ M$_{\sun}$.}
\label{app2}
\end{figure}

In this subsection, we study the numerical artifacts that could render our result non-physical. We have conservatively applied the \cite{2003MNRAS.338...14P} criterion to our DM haloes to estimate the radius within which the two-body relaxation time is shorter than the simulation time. We find a convergence radius of 73 pc for the M31 DM halo. The green vertical dashed line in Figures ~\ref{fgR5},~\ref{app1} and ~\ref{fgR6} marks this radius below which the simulations are not fully converged according to the criterion of \cite{2003MNRAS.338...14P}.

Next, we study the effect of the softening parameter, which models the interaction between two Plummer point masses. In order to test and assure that our simulation results are robust and reliable, we have performed softening tests. Figure~\ref{app1} shows the DM density profile for a set of simulations that only differ in the value of the softening length $\epsilon$. This figure demonstrates that doubling the softening radius leaves the density profile intact.

Furthermore, we study the effect of mass resolution and run our simulations for three different cases. We see in Figure~\ref{app2} that for a very large mass resolution of $4.4\times10^5$ M$_{\sun}$ the convergence radius of \cite{2003MNRAS.338...14P} shown by the vertical orange line moves to the right and hence the change in the density is only reliable over a very small range of radii. However, we see convergence between the mass resolution of $8.8\times10^4$ M$_{\sun}$ and $4.4\times10^4$ M$_{\sun}$.

We emphasize that the flattening of the density profile is observed here over more than a decade between 73 pc and 1 kpc. A few previous studies \citep[e.g.][]{2004MNRAS.349.1117B} using  simple  models of head-on collisions have concluded that the cusps remains cusps in mergers. However, such studies could not resolve scales below 1 kpc \citep[e.g. Figure 4 in][]{2004MNRAS.349.1117B} which is precisely the scales at which the cusp profile flattens. This study was also not carried out in the original work of \cite{2014MNRAS.442..160S} again because of lack of numerical resolution.

In the next subsection, we provide  fits to the newly flattened profiles and study the same migration and heating mechanism in other models of formation of GSS in M31.

\subsection{Model-independent  core-like dark matter profile in M31}

\begin{figure}
\centering
\includegraphics[width=0.47\textwidth]{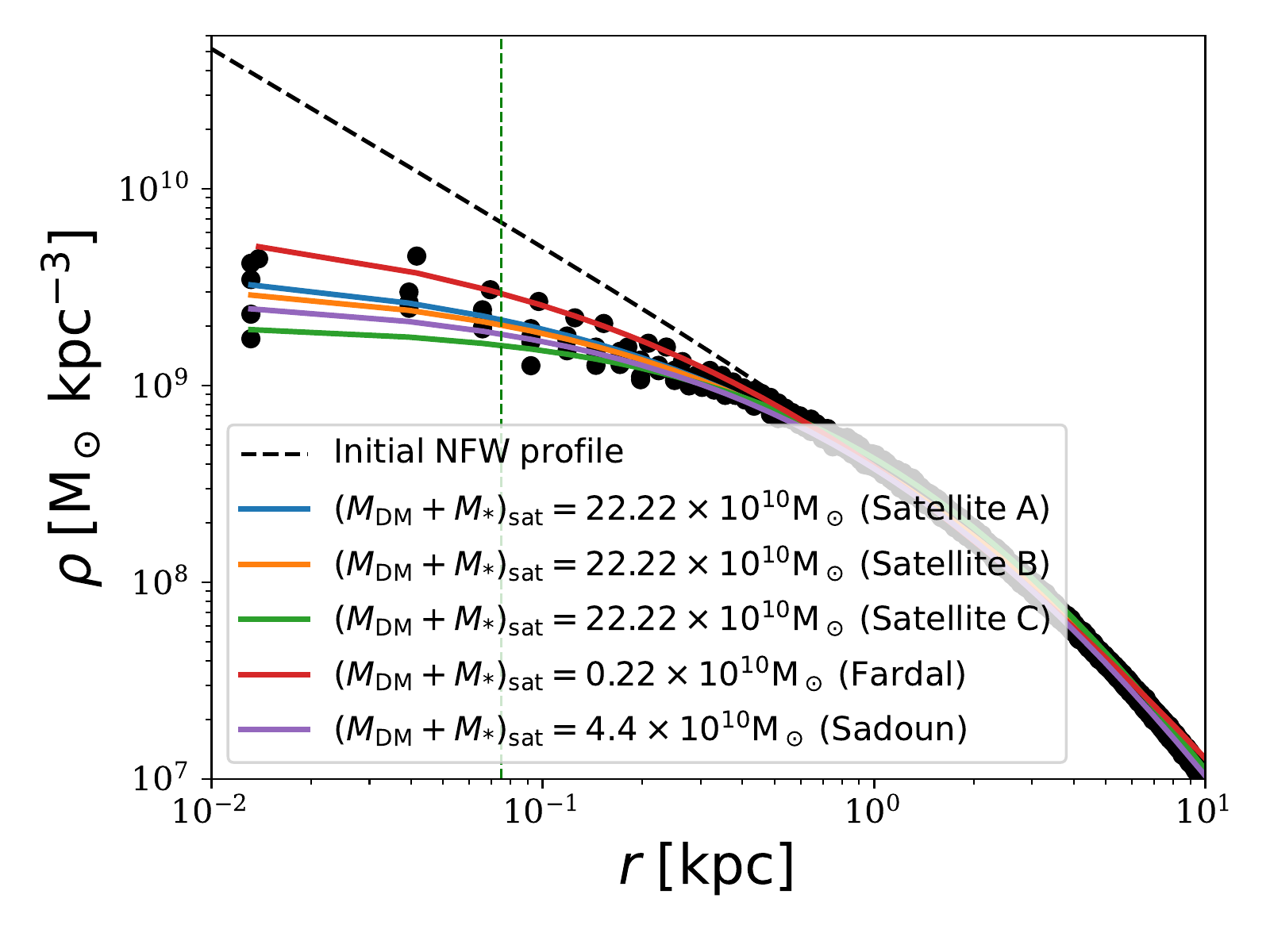}
\caption{{\it Model-independent core-like profile in M31:} Spherically-averaged DM density profile for all our five scenarios in 26 pc thick radial shells (see Table~\ref{tab1}). In all scenarios, the satellite has a stellar mass M$_{*}=2.2\times10^9$ M$_{\sun}$. Initially, the M31 DM halo assumes a NFW profile (black dashed line). We consider DM particles from both the M31 and satellite haloes to determine the DM density profile of M31. The fitting function described by Equation~\ref{eq1} reproduces the simulated density structures and captures the rapid transition from the cusp to the core. We set Poissonian errors for fitting weights. We stress that our best-fit core radii are larger than the numerical convergence radius marked by the vertical dashed green line. We observe a DM core of about 1.1 kpc for the M31 halo in our best-fit model \protect\citep{2014MNRAS.442..160S}.}
\label{fgR6}
\end{figure}

\begin{figure}
\centering
\includegraphics[width=0.47\textwidth]{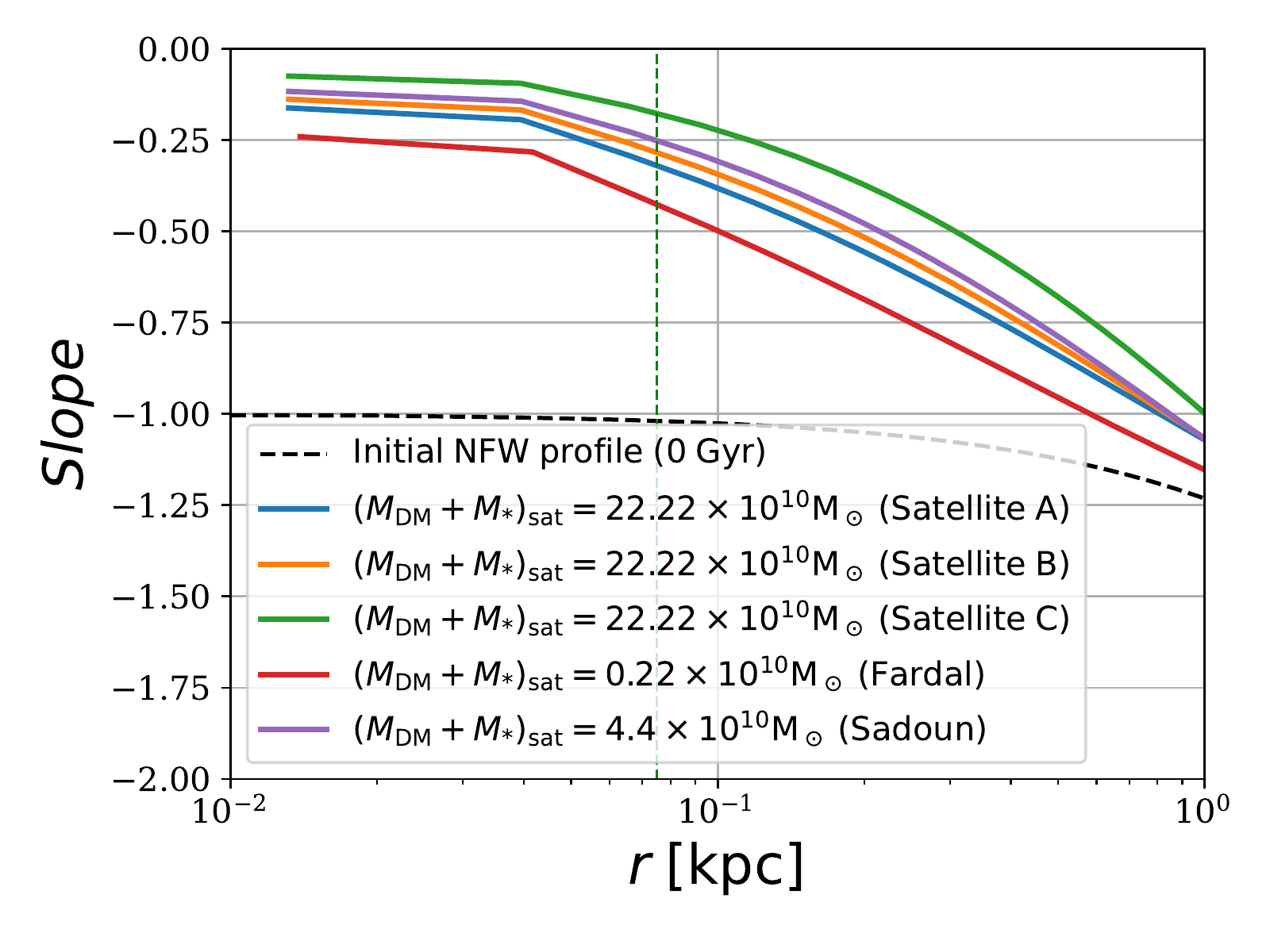}
\caption{{\it Core-like profiles of M31:} Inner slope of the DM density profile of M31 is plotted against the radii. In all scenarios, the satellite has a stellar mass M$_{*}=2.2\times10^9$ M$_{\sun}$. It clearly shows a flattening of the slope. For the Sadoun model and larger $M/M$, we clearly see a slope larger than -0.5 spanning over a wide range, which is often used as a definition of a core profile \citep{2001ApJ...552L..23D,2008AJ....136.2761O,2011AJ....141..193O,2015AJ....149..180O}.}
\label{fgR7}
\end{figure}

In order to affirm the model-independent nature of the cusp-to-core transition in M31, we have studied all of our five models (Table~\ref{tab1}).

We consider DM particles from both M31 and satellite haloes to determine the spherically averaged DM density profile for all our five models. As shown in Figure~\ref{fgR6}, we find that our profile is well-fitted by the following four-parameter formula: \citep{1993MNRAS.265..250D,1996MNRAS.278..488Z,1992MNRAS.254..132S,1990ApJ...356..359H}:
\begin{equation}
    \rho(r) = \frac{\rho_{\mathrm{c}}}{(1+(r/r_{\mathrm{c}})^{1/\beta})^{\gamma}},
\label{eq1}
\end{equation}
where $\rho_{\mathrm{c}}$ is the core constant density and $r_{\mathrm{c}}$ is the core radius which we find to be about 1.1 kpc for the M31 halo in the Sadoun scenario (see Table~\ref{tab1}). We stress that our best-fit core radii are larger than the numerical convergence radius (73 pc) and persist over almost one decade. However, as the value of the core size depends on the fitted DM profile, we defer from imposing the precise value of the radius as a constraint on the M31 core and here we only demonstrate that a merger with a wide range of halo-to-stellar mass ratio can indeed flatten the central density profile.

A central question in our work is whether or not the initial density profile has become so much shallower that it can  be called a core. There are various definitions for a core. The simplest evidence for a core is the visual change in the density profile. Figure~\ref{fgR6} clearly shows that the density profile above the convergence radius of 73 pc has been flattened and can no longer be fitted by the same NFW profile. A more in-depth evidence is given by the value of $r_{\mathrm{c}}$ in Equation~\ref{eq1} which  is larger than the convergence radius and defined by the radius at which the constant density is approximately divided by two. The third piece of evidence is provided by direct measurement of the slopes of the density profiles as shown in Figure~\ref{fgR6}. Different ranges of values of these slopes have been used to define a central core \citep{2001ApJ...552L..23D,2008AJ....136.2761O,2011AJ....141..193O,2015AJ....149..180O}. Figure~\ref{fgR6} clearly shows a flattening of the slope. For the Sadoun model and larger $M_{\mathrm{DM}}/M_{*}$, we clearly see a slope larger than $-0.5$ spanning over a wide range. Whether this flattening continues to smaller radii below our convergence radius is a question that will be pursued in forthcoming work where we will exploit more powerful GPU resources.

For the Fardal model \citep{2007MNRAS.380...15F}, the density profile at the end of the simulation , which is at 0.85 Gyr when the best-match with observations is obtained, in spite of the satellite being far less massive in this scenario, is well-fit by a core-like profile (see Figure~\ref{fgR6}). However, the core size is relatively smaller as is expected because the heating and migration mechanisms are less effective for the low-mass satellite. As the satellite in the Sadoun model is 20 times more massive than that in the Fardal model, the core size is larger in this scenario. 
In the other scenarios with a larger halo-to-stellar mass ratio, a cusp-to-core transition also occurs (see Figure~\ref{fgR6}).

It has been proposed that a major merger scenario could explain the kinematic data and stellar abundance data of M31 \citep{2010ApJ...725..542H,2018MNRAS.475.2754H,2018NatAs...2..737D}. Here, we demonstrate that a merger of satellites with $M_{\mathrm{DM}}/M_{*}=100$, which can be considered as major mergers, can also lead to the flattening of the initially cuspy profiles and yield new profiles, which are well-fitted by core-like functions (see Figure~\ref{fgR6}). Consequently, a major merger scenario with a satellite on a highly eccentric orbit should also trigger cusp-to-core formation for the M31 halo. Besides, the resulting DM core is expected to be larger as larger perturber masses lead to larger constant density central region \citep{2010ApJ...725.1707G}, which is shown by model C (green curve in Figure~\ref{fgR6}). Model B, which conforms best with observations for $M_{\mathrm{DM}}/M_{*}=100$, also shows a clear flattening of the DM density profile of M31 (see Figure~\ref{fgR6}). In spite of the fact that models A and C cannot reproduce the observational data on the GSS, they disturb the density profile of the host galaxy and flatten the central cusp, which hints that such a transition can be a generic outcome of most close-encounter merger scenarios.

In addition to the satellite mass and orbit, another dynamical parameter, which impacts the flattening of the cusp, is the concentration of the DM and star distributions within the satellite. Even if the satellite in models A, B and C is more massive than Fardal’s satellite, the satellite DM distribution in these three models is more diffuse compared to the one of Fardal, which is only formed by stars. Indeed, it is harder to slow down a satellite with an extended DM distribution as the mass loss of the satellite can easily compensate the dynamical friction from the DM background. Consequently, the energy transferred to the DM central region of M31 is reduced and this  explains why the DM density profiles only differ by a factor of 2 at the convergence radius between these models (see Figure~\ref{fgR6}). Even if dynamical friction increases in proportion to the satellite mass, the DM halo at large radius is not sufficiently dense compared to the central region to  significantly slow down the satellite. The transferred energy via dynamical friction by the satellite to the DM outer regions of M31 is found to be negligible as depicted by the DM density profiles in Figure~\ref{fgR6}.

\subsection{Comparison with cosmological simulations}

The fact that all of our models with $M_{\mathrm{DM}}/M_{*}$ ranging from 0 to 100, i.e. spanning minor to major mergers, redistribute DM in the central part of M31 and are well-fitted by a core-like profile is strong evidence for the universality of our result. We expect the flattening of the initial cuspy profile to occur {\it universally} at least in nearly radial close-encounter merger scenarios as discussed here. However, there is a widespread consensus that the hierarchical assembly of CDM halos yields a cuspy DM profile in DM-only simulations \citep{1996ApJ...462..563N,1997ApJ...490..493N,1999ApJ...524L..19M,2001ApJ...557..533F,2004MNRAS.349.1039N,2004MNRAS.353..624D}. These cosmological simulations find cuspy halos as we do when we run simulations with low mass resolution (see Figure~\ref{app2}). In such cases there is enough resolution to resolve the energy transfers via dynamical friction between particles in the central region of galaxies.  In the absence of particles in the central regions, the energy transfer induced by the dynamical friction is not properly modelled. The mass resolution is directly related to the convergence radius, below which we cannot resolve the physics in the simulation. As demonstrated by \cite{2010MNRAS.402...21N} and our Figure~\ref{app2}, it is necessary to have mass resolution lower than $10^5$ M$_{\sun}$ to investigate properly the physics below the kpc scale. For instance, in the  Illustris TNG simulations,  the central regions of galaxies are not resolved as the best resolution in these simulations is $4.5\times10^5$ M$_{\sun}$,  achieved in the latest TNG 50 simulation \citep{2019MNRAS.490.3196P}.

Indeed, the DM density profile of resimulated halos in zoom simulations reveals that the density profiles become shallower inwards down to the innermost resolved point \citep{2010MNRAS.402...21N}. Here we are demonstrating that the subhalo accretion could be at the origin of such flattenings of the NFW profile. Even if subhalos orbits become more radial and plunge deeper into their host halo at higher host halo mass \citep{2011MNRAS.412...49W}, the cuspy profile persists if the mass of the infalling stellite is at the lower end of the mass range. In our M31 scenario, however, the flattening of the cusp is more easily detected because the satellite is massive and falls in on a highly radial orbit.

\begin{figure}
\centering
\includegraphics[width=0.47\textwidth]{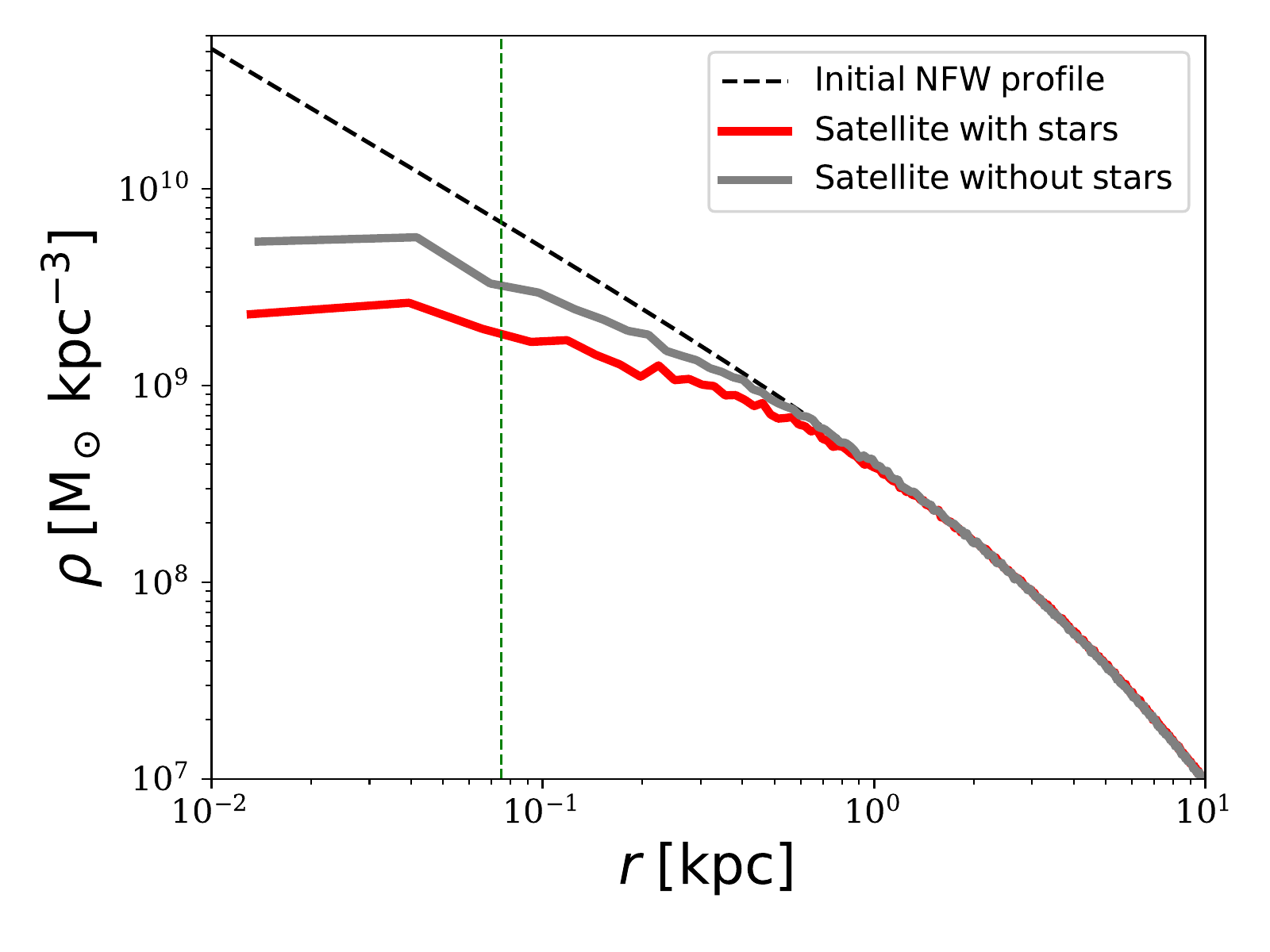}
\caption{{\it Impact of stars from the satellite:} DM density profile of M31 after the infall of satellite with (without) stars in 26 pc thick radial shells in the Sadoun scenario (see Table~\ref{tab1}). Initially, the M31 DM halo assumes a NFW profile (black dashed line). We stress that the stellar component of the satellite has a non-negligible contribution to the flattening of the DM cusp.}
\label{fgR8}
\end{figure}

Moreover, we highlight also that the stellar component of the satellite play a major role in core formation. Indeed, as this component is more concentrated compared to the dark matter of the satellite (see Table~\ref{tab1}), it will further slow down the satellite during its infall and thus disturb the central region of M31 even more prominently. Figure~\ref{fgR8} compares the DM density profile of M31 after the infall of a satellite with and without a stellar component in the Sadoun scenario (see Table~\ref{tab1}). The stellar component of the satellite has a non-negligible contribution to the flattening of the DM cusp and this clearly demonstrates that stellar component cannot be neglected in the cosmological simulations which aim at determining the  density profiles of dark matter haloes.

%--------
\section{Conclusion}
%--------

The giant stellar stream and the shell-like features of M31 are likely outcomes of accretion of  satellites on  highly eccentric orbits as shown by numerical simulations in the past and reaffirmed here. In this work, we have
mainly studied the model proposed by Sadoun in 2014 in which the Satellite is dark-matter rich and falls from its turnaround radius onto M31.

However, we have increased the resolution of the previous simulations by using a fully GPU code. The substantial gain of 100 in mass resolution has enabled us to here study the impact of such an accretion event on the spatial distribution of DM in the central regions of M31. Our simulations show that as the satellite falls onto M31, it is slowed down by dynamical friction and its energy is transferred to the host halo. The DM particles in the central regions of M31 are heated and migrate outwards. Here we have shown that in this process, the initial cusp becomes shallower   over almost a decade in radius and is  well-fitted by a core-like profile. 

To explore the model-independent aspects of our results, we have studied different  models with halo-to-stellar mass ratios ranging from 0 to 100 corresponding to minor and major merger scenarios. An ancient major merger scenario has been proposed for M31 \citep{2010ApJ...725..542H}, which however did not reproduce the observed features of the GSS. A new version of this model in which the merger occurs 2 Gyr ago (as in the Sadoun model) has been proposed recently \citep{2018MNRAS.475.2754H}, which has been taken further to also account for the origin of M32 \citep{2018NatAs...2..737D} but has also been proposed in the past \citep{2006Natur.443..832B}. Major mergers are quite rare, especially recently, and it has been shown that they could thicken the disk of galaxies by more than is allowed by observations. It is customary to study the stars from the merger but the more massive DM component could churn up the host galaxy and destroy any thin disk \citep[e.g.][]{1992ApJ...389....5T}.

We infer that merger events in which galaxies fall on highly eccentric orbits into their host haloes can provide a general mechanism for flattening of the density profile in a $\Lambda$CDM Universe where haloes are expected to be cuspy. It has been reported that satellites in host haloes with larger mass ratios have slightly more eccentric orbits with lower angular momentum, and moreover, satellites around more massive haloes seem to be on more radial orbits at fixed mass ratio \citep{1997MNRAS.290..411T,2011MNRAS.412...49W,2015MNRAS.448.1674J}. Hence, we expect  a noticeable fraction of galaxies in a $\Lambda$CDM Universe to harbour core-like profiles that have been formed during merger and accretion events. In this work, we have ignored the presence of a central black hole and have assumed an initially NFW profile for M31. The initially cuspy profile must have been even cuspier, due to the redistribution of DM around black holes \citep{1976ApJ...209..214B,1999PhRvL..83.1719G}. We however expect that the mechanism discussed here will not be influenced by a central black hole. A study of the fate of the black hole and the change of the central profile during the merger event studied here will be presented in  forthcoming work. Present observational data from stellar kinematic observations and the disc’s HI rotation curve have been inconclusive about the DM distribution in M31 \citep[e.g.][]{2012A&A...546A...4T,2018MNRAS.481.3210B}. Here, we have shown that GSS-constrained modelling of M31 favours a shallow cusp or a core-like inner DM distribution in M31. Future observations will shed light on this interesting question for our nearest neighbour.

Similarly, it was demonstrated that the recent passage of the Large Magelanic Cloud (LMC) has perturbed the MW disc \citep{2020NatAs.tmp..236P} and also distorted the structure of the DM halo \citep{2019MNRAS.488L..47B,2019ApJ...884...51G,2020MNRAS.498.5574E}. In contrast with our M31 merger scenario, the heating from the LMC seems to particularly affect the outer regions of the MW halo as the LMC has a much larger pericentre ($\sim$ 50 kpc) than our M31 satellite. With more powerful GPU resources, we could investigate the impact of the LMC infall on the inner regions of the DM halo of the MW.

\acknowledgments
We thank the reviewer for constructive feedback which helped to improve the quality of the manuscript. We thank Henry McCracken for constructive comments and discussions. We also thank Miki Yohei for providing us with the non-public $N$-body code, \textsc{gothic}.





%\bibliography{sample63}{}
%\bibliographystyle{aasjournal}

\bibliographystyle{aasjournal}



\end{document}